# Landau model for the phase diagrams of the orthorhombic rare-earth manganites RMnO$_3$

J. L. Ribeiro and L. G. Vieira

Centro de Física, Universidade do Minho, 4710-057 Braga, Portugal

The present work aims to describe, within a single phenomenological approach, the specific sequence of phase transitions observed in the rare-earth manganites RMnO$_3$ at zero magnetic field. It is shown that a single integrated description of the temperature versus composition phase diagrams of these compounds and related solid solutions can be obtained within the scope of Landau theory by adopting the so called type-II description of the modulated phases.

1. INTRODUCTION

The Landau theory of phase transitions constitutes a powerful tool to describe a great variety of phase transitions and, in particular, phase transitions involving modulated phases. The advantage of this phenomenological approach lies in its ability to establish a direct and exact relationship between the crystal symmetry and the physical properties of the system[1]. By establishing such a relationship, the theory allows us to describe the behaviour of macroscopic quantities (such as the polarization, magnetization, dielectric constant, etc) and to interpret the observed anisotropy or the relevant coupling mechanisms within one exact and symmetry based framework.

Over the last few years, a great deal of attention has been paid to several metal compounds in which ferroelectricity is induced by a transition to a complex magnetic state[2-10]. In this class of systems, external magnetic fields or chemical pressure fields originated from the partial substitution of a molecular unit, are capable of rotating or stabilizeing an electrical polarization.[11-17] Although these effects might mimic single phase effects like ferromagnetoelectricity (the linear magneto-electric effect) or piezoelectricity, they result from rather different mechanisms. Here, the magnetic or the stress fields induce magnetic phase transitions which, in turn, alter the symmetry of the system and modify the set of compatible secondary order parameters. It is from this modification of the symmetry that the change of the polar state of the system originates. Therefore, in this class of compounds, the remarkable cross effects between magnetic ordering and electric polarization relate more to the field of the critical phenomena and improper ferroelectricity than to the fields of the ferromagnetoelectricity or multi-ferroicity.

Among this novel class of improper ferroelectrics, the orthorhombic manganites $RMnO_3$ (R= Eu, Gd, Tb, Dy, Ho) and related solid solutions such as $Eu_{1-x}Y_xMnO_3$, $Gd_{1-x}Tb_xMnO_3$ or $Dy_{1-x}Tb_xMnO_3$ are those possessing the simplest crystallographic and magnetic structures. Because of this reason, these compounds constitute adequate model systems in which symmetry based models can be explored without the need of the full apparatus of group theory. The present work takes advantage of this fact to obtain, for these compounds, a single integrated model capable of accounting for the observed sequence of phase transitions. As it will be shown, this integrated picture can be obtained within the scope of the so called type-II Landau description of the modulated phases[18] and can be used to interpret, model and organize the experimental data concerning the temperature versus composition phase diagrams of the pure compounds and of their solid solutions.

2. THE CRITICAL BEHAVIOUR OF THE DIFFERENT $RMnO_3$ COMPOUNDS

At room temperature, the symmetry of the orthorhombic rare-earth manganites $RMnO_3$ is described by the paramagnetic group $G=(Pnma)´$ and the unit cell possesses four molecular formulae (Z=4).[19] The magnetic phases observed at lower temperatures result essentially from the ordering of the Mn spins $\vec{S}_1, \vec{S}_2, \vec{S}_3$ and $\vec{S}_4$ which, in the paramagnetic phase, are located in the unit cell at the positions (o,o,½), (½,o,o), (o,½, ½) and (½, ½, o), respectively[1]. At the centre of the Brillouin zone ($\vec{k}=0$), the 12 components of these spins generate a reducible co-representation $\Gamma$ of the paramagnetic space group, whose decomposition into a direct sum of irreducible co-representations leads to $\Gamma = 3A_g^- \oplus 3B_{g1}^- \oplus 3B_{g2}^- \oplus 3B_{g3}^-$. Here, the (-) superscript signals the odd character of these co-representations under time reversal. To each irreducible co-representation there will correspond a set of magnetic eigenvectors, as specified in table I.

---

[1] The spins of the rare-earth ion will also play a role, especially at lower temperatures but we will ignore their contribution because they are not essential for the understanding of the global phase diagrams observed.

TABLE I. Possible magnetic eigen-modes originated from the $Mn^{3+}$ located in 4b Wyckoff positions

| $A_g^-(\Delta_1)$ | $B_{2g}^-(\Delta_2)$ | $B_{3g}^-(\Delta_3)$ | $B_{1g}^-(\Delta_4)$ |
|---|---|---|---|
| $G_x$ | $F_y$ | $F_x$ | $F_z$ |
| $A_z$ | $A_x$ | $A_y$ | $G_y$ |
| $C_y$ | $G_z$ | $C_z$ | $C_x$ |

In this table, the eigen-vectors are denoted as $\vec{A} = \vec{S}_1 + \vec{S}_2 - \vec{S}_3 - \vec{S}_4$, $\vec{G} = \vec{S}_1 - \vec{S}_2 - \vec{S}_3 + \vec{S}_4$, $\vec{F} = \vec{S}_1 + \vec{S}_2 + \vec{S}_3 + \vec{S}_4$ and $\vec{C} = \vec{S}_1 - \vec{S}_2 + \vec{S}_3 - \vec{S}_4$. This notation directly specifies the relative orientation of the different spins. For example, for the mode $\vec{A}$, the spins pairs ($\vec{S}_1, \vec{S}_2$) and ($\vec{S}_3, \vec{S}_4$) are oriented parallel to each other within each set and both sets are anti-ferromagnetically coupled.

As in the prototype case of $LaMnO_3$ (Ref. 19), the $Mn^{3+}$ electronic configuration is $t_{2g}^3 e_g^1$, with the spin quantum number $S=2$. The three $t_{2g}^3$ electrons are localized, while the $e_g$ electron orbitals are extended in the basal (010) plane[2] and are strongly hybridized with the oxygen $p$ orbitals. Consequently, the ferromagnetic superexchange interactions of the $e_g^1$ electrons in the plane and the antiferromagnetic interactions of the $t_{2g}^3$ electrons out of the plane favour the onset, at low temperatures, of one antiferromagnetic order of the *A*-type (A-AFM). This is well apparent in the set of systems ranging from La to Sm, where a direct transition from the paramagnetic phase (PM) to one A-AFM phase is observed. Here, the AFM order results precisely from the stabilization of the $A_x$ irreducible magnetic mode (see table I). Consequently, the system acquires a symmetry described by the magnetic space group $Pnma(P1\frac{2_1}{m}1)$ [3] and a canted ferromagnetic moment directed along the $\vec{b}$-axis.

---

[2] We will adopt the Pnma setting, as opposed to the Pbnm setting used by some authors.

However, smaller rare-earth ions ($Eu^{3+}$, $Gd^{3+}$ $Tb^{3+}$, $Y^{3+}$ etc.) fit worse in the perovskite network and give rise to more pronounced *b*-axis rotations of the Mn-O octahedra. Consequently, the Mn-O-Mn bonding angles diminish and the orthorhombic distortion of the lattice increases as the ionic radius is reduced.[20] This effect weakens the in-plane ferromagnetic superexchange interactions and modifies the orbital overlap and the relative strength of the anti-ferromagnetic interaction between the next-nearest neighbours (NNN) $Mn^{3+}$ spins. The ferromagnetic order of the in-plane spins tends to become strongly frustrated.

At first, this geometrically driven effect simply decreases the Néel temperature from 140K (La) to 60K (Sm). However, beyond a certain threshold and within the range delimited by Eu and Ho, the magnetic instability shifts from the centre to the interior of the Brillouin zone (along the $\Sigma$-line), giving rise to an intermediate longitudinal incommensurate (L-INC) phase, with a modulation wave vector $\vec{k} = \delta(T)\vec{a}^*$ (Ref 20-23). The corresponding incommensurate order parameter is still irreducible and of the symmetry $\Gamma(B_2)$ (Ref. 24-27; see notation in Ref. 28), containing therefore the active mode $A_x$ in the limit $\vec{k} \to 0$ [$\lim_{\vec{k}\to 0}\Gamma(B_2) = B_{2g}^- \oplus B_{1u}^+$]. Given the one-to-one relationship existing in this case between each of the irreducible co-representations of the paramagnetic group and the symmetry of the corresponding modulated phase[28], this L-INC phase must have the symmetry described by the magnetic superspace group $P_a(P_{\bar{1}1S}^{nma})$ [4] (Ref. 29). This symmetry is incompatible with any ferroic or linear magneto-electric properties.

---

[3] Here, the magnetic symmetry is described by the unitary space group $P1\frac{2_1}{m}1$ plus non unitary operations ($\theta C_{2x}, \theta C_{2z}, \theta\sigma_x$ and $\theta\sigma_z$) that recover a $Pnma$ symmetry.

Figure 1 shows the value of the modulation wavenumber observed immediately below the transition from the PM phase, $\delta(T_i)$, as a function of the ionic radius $R_{ion}$ of the rare-earth element (data taken from Ref. 20). As seen, the L-INC phase sets in at a location in the Brillouin zone that varies linearly with $R_{ion}$. Therefore, the instability of the PM phase moves along a single magnon branch, approaching the Brillouin zone boundary (X- point, $\delta = 1/2$) as $R_{ion}$ decreases (see also Ref. 30). This drift of $\delta(T_i)$ can be extrapolated in the direction of larger or smaller values of $R_{ion}$. This extrapolation indicates that the centre and the border of the Brillouin zone would be reached for $R_{ion}$ of the order of 113pm and 102pm, respectively. This latter value is very close to the ionic radius of both $Yb^{3+}$ and $Lu^{3+}$ (100.8 pm and 100.1 pm, respectively), for which the orthorhombic $RMnO_3$ system shows direct transitions from the paramagnetic phase to one E-type antiferromagnetic (E-AFM) phase[5] (Ref. 30-34). The observed E-AFM order consists of [101] rows of spins coupling parallel to a neighbouring row on one side and antiparallel on the other with an antiparallel coupling between (010) planes. This spin structure corresponds to the one expected for the lock-in of one order parameter of symmetry $\Gamma(B_2)$ at the Brillouin zone edge $\vec{k} = \frac{1}{2}\vec{a}^*$ (Ref. 33), for a global phase $\Phi = \pi/4$ (Ref. 28). The corresponding magnetic group is $P_a(Pnm2_1)$ and a ferroelectric polarization directed along the c-axis is then allowed by symmetry.[28] This polarization is actually observed in the E-AFM phase of $HoMnO_3$, $YbMnO_3$ or $LuMnO_3$ (Ref. 16, 33, 34) (as well as in the E-AFM phase of some nickelates[35-38]). It provides one striking example of improper ferroelectricity driven by an irreducible magnetic order parameter. It cannot be explained by a Dzyaloshinskii-Moriya mechanism,[39,40] the spin current model[41], the electric current cancellation model[42] or by the usual heuristic pictures.[43] The above observations can be summarized by saying that, over the whole set of orthomanganese compounds, the primary instability of the paramagnetic phase moves from the centre to the edge of the Brillouin zone along a single magnon branch of symmetry $\Gamma(B_2)$. This common symmetry of the primary order parameter will be essential for our present purposes. However, one additional question concerning the possible role of other (secondary) magnetic distortions must be analysed in order to elucidate the relevant set of magnetic order parameters. Let us first notice that, in systems such as TbMnO$_3$ or DyMnO$_3$, there occur, at lower temperatures, cycloidal phases resulting from the stabilization of order parameters of symmetry $\Gamma(B_2)+\Gamma(A_2)$ (at zero magnetic field) or, in the case of TbMnO$_3$, $\Gamma(B_2)+\Gamma(A_1)$ under magnetic fields.[25,44] Besides the primary branch $\Gamma(B_2)$, the additional branches involved here, have symmetries $\Gamma(A_2)$ and $\Gamma(A_1)$. These branches correspond to spatial modulations of the magnetic modes $A_y$ and $A_z$ given in table I [$\lim_{\vec{k}\to 0}\Gamma(A_2)=B_{3g}^{-}\oplus A_u^{+}$ and $\lim_{\vec{k}\to 0}\Gamma(A_1)=A_g^{-}\oplus B_{3u}^{+}$]. Also, in HoMnO$_3$, for example, the observed diffraction patterns correspond to parent reflections (*hkl*) satisfying *h+l=2n* and *k=2n+1* (*n* integer).[30] Given that the Mn$^{3+}$ ions occupy, in the paramagnetic unit cell, 4b Wyckoff sites, the observed reflection conditions imply that the spin wave can only include *A*-modes. Hence, even if the $A_x$ mode may be seen as primary, all the three A-modes seem to play a role in the phase transition sequences observed in the RMnO$_3$ system. Other non-magnetic secondary order parameters, that are allowed by the symmetry, must also be taken into account.

---

[4] $P_a(P_{\bar{1}1S}^{nma})$ denotes $P_{\bar{1}1S}^{nma} \otimes \left[\{E;000,0\},\{\theta;000,1/2\}\right]$, where $P_{\bar{1}1S}^{nma}$ is the unitary superspace group (in the standard notation).
[5] For Ho-Lu, Y and Sc, the lower energy structure becomes a layer like structure with hexagonal symmetry ($P6_3cm$). However, even in this range of smaller ions, high pressure techniques, low soft chemistry or epitaxial thin film growth still allow the synthesis of orthorhombic compounds.

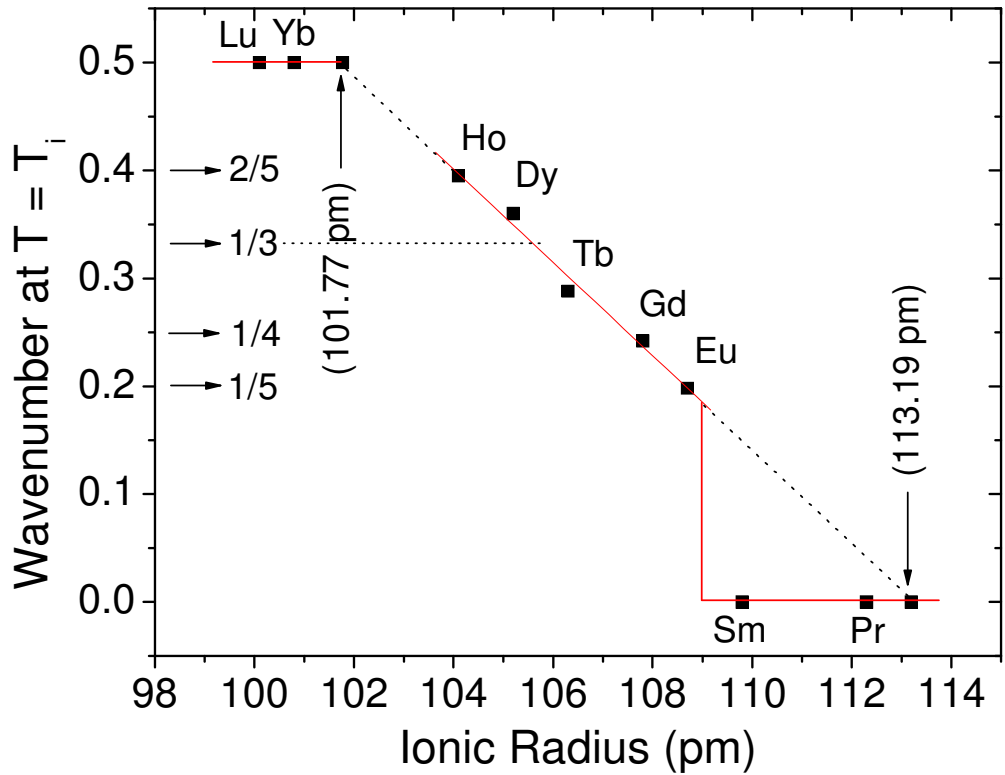

FIG. 1. The ionic radius of the rare-earth elements (from Pr to Lu) plotted as a function of the modulation wavenumber immediately below the stability limit of the paramagnetic phase. The straight line from Ho to Eu is a linear fit. The points marked by arrows at the ionic radii 101.77 pm and 113.19 were obtained by extrapolation of that best fit.

## 3. THE LANDAU FREE ENERGY

In the usual Landau theoretical framework for incommensurate systems, both the lock-in commensurate phase (either homogeneous or modulated) and the incommensurate phase are described by a common primary order parameter: the symmetry of the primary distortion is kept invariant while the modulation wavevector changes with temperature, composition or external fields.[1] In the case of the orthorhombic rare-earth manganites RMnO$_3$, the modulation wave vector is kept fixed along a main crystallographic direction ($\vec{k} = \delta \vec{a}^*$) and, as seen in the preceding section, the primary order parameter maintains its symmetry $\Gamma(B_2)$, over the entire range of the rare-earth elements. It is this common symmetry of the primary parameter over the whole set of compounds that allows us to deal with the different systems within one unified model.

One incommensurate phase is normally seen as a modulation or a periodic distortion of a given underlying commensurate or lock-in phase. The dimension of the incommensurate order parameter will therefore depend on the dimension of the basic commensurate order parameter chosen for the description. For the case in hand, if the lock-in wave-vector is located inside the Brillouin zone, then the commensurate order parameter is a complex number representing the amplitude of the magnetic wave and its global phase with respect to the underlying lattice. The incommensurate modulation (the distortion of the lock-in phase) is here stabilized by a Lifshitz invariant[45] and the evolution of the wavenumber towards its lock-in value originates from the competition between this term and the Umklapp potentials favouring the commensurate order. This case corresponds to the so called type-I description of an incommensurate phase.[46] If, on the other hand, the lock-in phase is taken as homogeneous ($\vec{k} = 0$) then, for the symmetry here considered, the order parameter is necessarily one-dimensional and the Lifshitz invariant forbidden. Then, the stabilization of a modulated spin structure may only be obtained by considering a free energy density expansion containing invariants which depend on the spatial derivatives of the one-dimensional primary order parameter. This second description of a modulated incommensurate phase is called type-II[46]. It is far more versatile if one wants to go beyond the description of a particular phase transition and capture, within a single phenomenological model, a sequence of phase transitions, a temperature versus magnetic field or a temperature versus composition phase diagram involving a single critical branch. This versatility has been well demonstrated in the case of displacive systems like sodium nitrite,[47] thiourea[48] or BCCD[49] and, with the necessary adaptations, can be used for the case of the RMnO$_3$ compounds.

### A. The Landau free energy density

Let us then consider the problem of finding the adequate type-II free energy density expansion. For the chosen lock-in wave vector ($\vec{k} = 0$), the primary order parameter is the $A_x$ magnetic mode. In order to accommodate the possible stabilization of a modulated spin structure, the free energy density must include terms depending on the spatial derivatives of this mode. The symmetry constrains that are verified here correspond exactly to those that are observed in displacive systems like NaNO$_2$, SC(NH$_2$)$_2$ or BCCD.[47-49] Consequently, similar to these systems, this part of the free energy density can be written as:

$$f_1 = \frac{1}{2}\alpha_x A_x^2 + \frac{1}{4}\beta_x A_x^4 - \frac{1}{2}\sigma(\frac{\partial A_x}{\partial X})^2 + \frac{1}{4}\gamma(\frac{\partial^2 A_x}{\partial X^2})^2 + \nu A_x^2 (\frac{\partial A_x}{\partial X})^2 \qquad (1)$$

The dispersive term $\nu A_x^2 (\frac{\partial A_x}{\partial X})^2$ is symmetry allowed because it is the product of two trivial invariants. It is this term that imposes the temperature dependence of the modulation wave vector, favouring energetically smaller or higher values of the modulation wave vector if $\nu > 0$ or $\nu < 0$, respectively. Negative ($-\sigma$) and positive ($\gamma$) coefficients must be chosen in order to stabilize a minimum in the dispersion of the quadratic term at an arbitrary point of the Brillouin zone, a necessary condition for the occurrence of modulated spin structures. As usual, we will take $\alpha_x = \alpha_{x0}(T-T_0)$ with $\alpha_{0x} > 0$ and $\beta_x > 0$.

In the case of the secondary magnetic order parameters $A_y$ and $A_z$, we will adopt the simplest possible free energy, limiting the expansion to terms up to the fourth order. Also, the bi-quadratic mixed terms $A_x^2 A_y^2$ and $A_x^2 A_z^2$ will be considered to describe the coupling between the primary and the secondary magnetic order parameters. Consequently, we have:

$$f_2 = \frac{1}{2}\alpha_y A_y^2 + \frac{1}{4}\beta_y A_y^4 + \frac{1}{2}\alpha_z A_z^2 + \frac{1}{4}\beta_z A_z^4 + \Delta_1 A_x^2 A_y^2 + \eta_1 A_x^2 A_z^2 \qquad (2)$$

Here, we will take $\beta_y, \beta_z, \Delta_1, \eta_1 > 0$ and $\alpha_{y(z)} = \alpha_{y(z)0}(T - T_{1(2)})$, with $T_1 < T_0$ and $T_2 < T_1$. This latter choice means that we will assume that the secondary magnetic order parameters also possess intrinsic instabilities, although at temperatures lower than $T_0$. Hence, in the absence of any interaction between the three magnetic order parameters (that is, when $\Delta_1 = \eta_1 = 0$), a sequence of second order transitions would occur between phases characterized by the order parameters $A_x$, $A_x \oplus A_y$ (or $A_x \oplus A_z$) and $A_x \oplus A_y \oplus A_z$. The positive sign chosen for $\Delta_1$ and $\eta_1$, however, implies the possible suppression of this phase sequence and the first order character of an eventual transition between any two of these magnetic phases. Notice that for $\Delta_1 < 0$ and/or $\eta_1 < 0$, trigger-type phase transitions could occur even without any intrinsic instability of the secondary magnetic order parameters.

In addition to the pure magnetic invariants considered so far, we must also take into account other terms coupling the magnetic degrees of freedom with other secondary parameters. Here, because we are interested in the possibility of improper

TABLE II. Transformation properties of the terms bilinear in the magnetic order parameters or involving the first order spatial derivative of the primary order parameter.

| | $C_{2x}$ | $C_{2y}$ | $i$ | $\theta$ | |
|---|---|---|---|---|---|
| $x$ | 1 | -1 | -1 | 1 | $B_{3u}^{+}$ |
| $A_x$ | -1 | 1 | 1 | -1 | $B_{2g}^{-}$ |
| $A_y$ | 1 | -1 | 1 | -1 | $B_{3g}^{-}$ |
| $A_z$ | 1 | 1 | 1 | -1 | $A_{g}^{-}$ |
| $A_y \dfrac{dA_x}{dx}$ | -1 | 1 | -1 | 1 | $B_{2u}^{+}$ |
| $A_y A_x$ | -1 | -1 | 1 | 1 | $B_{1g}^{+}$ |
| $A_z \dfrac{dA_x}{dx}$ | -1 | -1 | -1 | 1 | $B_{3u}^{+}$ |
| $A_z A_x$ | -1 | 1 | 1 | 1 | $B_{2g}^{+}$ |

ferroelectricity or improper ferroelasticity, we will consider the particular case of coupling terms that are linear on one electric polarization or on one homogeneous lattice strain. From the transformation properties of the first spatial derivative of the primary parameter, $\dfrac{\partial A_x}{\partial x}$ and of products such as $A_y \dfrac{dA_x}{dx}$, $A_z \dfrac{dA_x}{dx}$, $A_y A_x$ and $A_y A_x$ (see table II), it becomes clear that the polarizations $P_y$ and $P_z$, [which are transformed, under (Pnma)´, as $B_{2u}^{+}$ and $B_{3u}^{+}$, respectively], along with the lattice deformations $e_{xy}$ and $e_{xz}$ [$B_{1g}^{+}$ and $B_{2g}^{+}$, respectively] are potential secondary parameters allowed by symmetry. The contribution of these mixed terms to the free energy density is of the form [6]:

---

[6] If one considers an external magnetic field B, additional terms $\left[\delta A_x^2 + \lambda_1 A_y^2 + \lambda_2 A_z^2 + (2\chi_m)^{-1}\right]M^2 - MB$ must be considered (here M is the induced magnetization). Here, however, we will consider only the case B=0.

$$f_3 = \Delta_2 A_y \left(\frac{\partial A_x}{\partial X}\right) P_y + \Delta_3 A_y A_x e_{xy} + \eta_2 A_z \left(\frac{\partial A_x}{\partial X}\right) P_z + \eta_3 A_z A_x e_{xz} +$$

$$+ \frac{P_y^2}{2\chi_y} + \frac{P_z^2}{2\chi_z} + \frac{e_{xy}^2}{2c_{xy}} + \frac{e_{xz}^2}{2c_{xz}} \tag{3}$$

Notice that, for simplicity, we have neglected invariants involving more than two secondary parameters, like $P_x P_y e_{xy}$, $P_x P_z e_{xz}$ or $A_y A_z e_{yz}$. This means that we are neglecting the potential stabilization of phases with very low symmetry. Although these phases may play a role in the detailed mechanisms for a given phase transition,[29,50,51] they are not essential for the global picture we pursue here. Also, we are ignoring the eventual commensurate character of the spin wave by not including eventual mixed Umklapp terms that are allowed for particular types of commensurate phases.[28,29]

The free energy density $f = f_1 + f_2 + f_3$ corresponds to the simplest possible functional with the potential to describe the observed zero field phase diagrams of the RMnO$_3$ compounds. However, it still contains an undesirable large number of adjustable constants.

### B. Reduced variables and some simplifying assumptions

As usual, the first step to improve the situation regarding the number of model parameters is the elimination of a number of physically irrelevant coupling constants. This can be achieved by expressing the free energy density in terms of dimensionless quantities. By defining $g = \frac{\gamma^2 \beta_x}{16\sigma^4} f$, $A_i = \frac{2\sigma}{(\gamma\beta_x)^{1/2}} S_i$, $X = (\frac{\gamma}{2\sigma})^{1/2} x$,

$\frac{\beta_x^{1/2} \gamma}{2\sigma^2 \chi_y^{1/2}} P_{y(z)} = p_{y(z)}$, $a_{y(z)} = \frac{\alpha_{y0}(\alpha_{z0})}{\alpha_{x0}}$, $b_{y(z)} = \frac{\beta_y(\beta_z)}{\beta_x}$, $\bar{\chi}_z = \frac{\chi_z}{\chi_y}$,

$t = \frac{\gamma}{8\sigma^2} \alpha_{0x}(T - T_0)$, $\beta_x^{-1}\Delta_1(\eta_1) = \nabla_1(\xi_1)$, $\nabla_2(\xi_2) = \Delta_2(\eta_2)\left[\frac{\chi_y \sigma}{2\beta_x \gamma}\right]^{1/2}$ and

$\nabla_3(\xi_3) = \frac{\gamma}{4\sigma^2}\Delta_3(\eta_3)$, one obtains the simpler reduced free energy density:

$$g(x) = tS_x^2 + \frac{1}{4}S_x^4 - \frac{1}{4}\left(\frac{\partial S_x}{\partial x}\right)^2 + \frac{1}{4}\left(\frac{\partial^2 S_x}{\partial x^2}\right)^2 + \mu S_x^2\left(\frac{\partial S_x}{\partial x}\right)^2 + a_y(t-t_1)S_y^2 + \frac{1}{4}b_y S_y^4 +$$

$$+ \nabla_1 S_x^2 S_y^2 + \nabla_2 p_y S_y (\frac{\partial S_x}{\partial x}) + \nabla_3 S_x S_y e_{xy} + a_z(t-t_2)S_z^2 + \frac{1}{4}b_z S_z^4 + \xi_1 S_x^2 S_z^2 +$$

$$\xi_2 p_z S_z (\frac{\partial S_x}{\partial x}) + \xi_3 S_x S_z e_{xz} + \frac{p_y^2}{2} + \frac{p_z^2}{2\bar{\chi}_z} + \frac{e_{xy}^2}{2\bar{c}_{xy}} + \frac{e_{xz}^2}{2\bar{c}_{xz}} \quad (4)$$

Moreover, we will introduce two additional approximations that will allow us to reduce further the number of the parameters and will help us to simplify the calculations to be made. Firstly, we notice that, for a cubic perovskite, $\bar{\chi}_z = a_y = a_z = b_y = b_z = 1$. Although this is no longer true in the presence of an orthorhombic distortion, it seems reasonable to assume that one can keep these values as a first approximation and eliminate five non-essential adjustable parameters. We will also use a simple plane wave to describe the magnetic modulation induced by the primary order parameter $S_x$. This second approximation can be justified by noticing that, over the whole temperature range of stability of the observed longitudinal or cycloidal modulated phases, essentially only first order satellites are observed in neutron or x-ray measurements. The plane wave approximation is therefore expected to describe reasonably well the magnetic modulation over the whole temperature and magnetic field ranges explored experimentally. Accordingly, we will write in (4) $S_x = \sigma_x \cos(qx)$, obtaining:

$$g(x) = \left[t\cos^2(qx) - \frac{q^2}{4}\sin^2(qx) + \frac{q^4}{4}\cos^2(qx)\right]\sigma_x^2 + \frac{1}{4}\left[\cos^4(qx) + 4\mu q^2 \sin^2(qx)\cos^2(qx)\right]\sigma_x^4 +$$

$$+ \frac{1}{2}(t-t_1)S_y^2 + \frac{1}{4}S_y^4 + \nabla_1 S_y^2 \sigma_x^2 \cos^2(qx) - \nabla_2 S_y \sigma_x p_y \sin(qx) + \nabla_3 S_y \sigma_x e_{xy} \cos(qx) +$$

$$+ \frac{1}{2}(t-t_2)S_z^2 + \frac{1}{4}S_z^4 + \xi_1 S_z^2 \sigma_x^2 \cos^2(qx) - \xi_2 S_z \sigma_x p_z \sin(qx) + \xi_3 S_z \sigma_x e_{xz} \cos(qx) +$$

$$+ \frac{p_y^2}{2} + \frac{p_z^2}{2\bar{\chi}_z} + \frac{e_{xy}^2}{2\bar{c}_{xy}} + \frac{e_{xz}^2}{2\bar{c}_{xz}} \quad (5)$$

## C. The non-magnetic order parameters

The equilibrium value of a given secondary and non-magnetic order parameter $X$ (here, as seen, $X = P_y, P_z, e_{xz}$ or $e_{xy}$) can be determined by imposing in (5) the condition $\frac{\partial g}{\partial X} = 0$. This leads to the following relations between non-magnetic and magnetic order parameters:

$$\begin{aligned} p_y &= \nabla_2 q \sigma_x S_y \sin(qx) \\ p_z &= \xi_2 q \sigma_x S_z \sin(qx) \\ e_{xy} &= -\bar{c}_{xy} \nabla_3 \sigma_x S_y \cos(qx) \\ e_{xz} &= -\bar{c}_{xz} \xi_3 \sigma_x S_z \cos(qx) \end{aligned} \qquad (6)$$

Notice that, here, the improper polarizations $p_z$ and $p_y$ can only occur in magnetic modulated phases ($q \neq 0$) involving at least two irreducible components of the magnetic modulation (note again that we are ignoring the eventual commensurate nature of the modulation wave vector), while the lattice deformations can be maintained even in the case of a homogeneous phase. By substituting (6) into (5) one can then express the free energy density as a function of the magnetic order parameters:

$$\begin{aligned} g(x) &= \left[ t \cos^2(qx) - \frac{q^2}{4} \sin^2(qx) + \frac{q^4}{4} \cos^2(qx) \right] \sigma_x^2 + \frac{1}{4} \left[ \cos^4(qx) + 4\mu q^2 \sin^2(qx) \cos^2(qx) \right] \sigma_x^4 + \\ &\quad + \frac{1}{2}(t - t_1) S_y^2 + \frac{1}{4} S_y^4 + \frac{1}{2} g S_y^2 \sigma_x^2 \cos^2(qx) - \frac{1}{2} \nabla_2^2 q^2 S_y^2 \sigma_x^2 \sin^2(qx) + \\ &\quad + \frac{1}{2}(t - t_2) S_z^2 + \frac{1}{4} S_z^4 + \frac{1}{2} h S_z^2 \sigma_x^2 \cos^2(qx) - \frac{1}{2} \xi_2^2 q^2 S_z^2 \sigma_x^2 \sin^2(qx) \end{aligned} \qquad (7)$$

Here, we have defined $g = 2\nabla_1 - \bar{c}_{xy} \nabla_3^2$ and $h = 2\xi_1 - \bar{c}_{xz} \xi_3^2$.

In the free energy density (7), the order parameter $A_x$ plays a central role not only because it softens at a higher temperature but also because it is the one that gives rise to the spin modulation wave. It is the instability of this primary mode that can trigger the stabilization of a modulation wave of $S_y$ or $S_z$, whose intrinsic instabilities would otherwise give rise to homogeneous phases. In fact, by imposing in (7) the conditions $\frac{\partial g(x)}{\partial S_y} = 0$ and $\frac{\partial g(x)}{\partial S_z} = 0$ one readily obtains:

$$S_y^2 = \sigma_{yF}^2 \cos^2(qx) + \sigma_{yA}^2 \sin^2(qx) \quad (8a)$$
$$S_z^2 = \sigma_{zF}^2 \cos^2(qx) + \sigma_{zA}^2 \sin^2(qx) \quad (8b),$$

with $\sigma_{yF}^2 = -[(t-t_1) + g\sigma_x^2]$, $\sigma_{yA}^2 = -[(t-t_1) - \nabla_2^2 q^2 \sigma_x^2]$, $\sigma_{zF}^2 = -[(t-t_2) + h\sigma_x^2]$ and $\sigma_{zA}^2 = -[(t-t_2) - \xi_2^2 q^2 \sigma_x^2]$.

As seen, it is the onset of a longitudinal modulation wave $S_x = \sigma_x \cos(qx)$ that can provoke, via the bi-quadratic coupling terms between the magnetic parameters, to the modulation wave of the components of $S_y$ or $S_z$. It is also clear that these secondary magnetic modulations can be mainly established either in phase ($\sigma_{yF}$ or $\sigma_{zF}$) or in quadrature ($\sigma_{yA}$ or $\sigma_{zA}$) with respect to $S_x$, depending on the values of the coefficients. These two possible sets of components of the secondary magnetic modulation have in fact different symmetries and couple, consequently, to different homogeneous parameters. For example, as we have seen in (6), $p_y = \nabla_2 q S_y \sigma_x \sin(qx)$ and $e_{xy} = -\bar{c}_{xy} \nabla_3 S_y \sigma_x \cos(qx)$. If, for example, $S_y = \sigma_{yF} \cos(qx)$, then $p_y = \nabla_2 q \sigma_{yF} \sigma_x \sin(qx)\cos(qx)$ and $e_{xy} = -\bar{c}_{xy} \nabla_3 \sigma_{yF} \sigma_x \cos^2(qx)$. That is, if $S_y$ is in phase with $S_x$, the value of the polarization wave, averaged over one period of the modulation, is null, $\langle p_y \rangle = 0$, while the lattice deformation $e_{xy}$ is not ($\langle e_{xy} \rangle \neq 0$). Conversely, if $S_y = \sigma_{yA} \sin(qx)$, then $\langle p_y \rangle \neq 0$ and $\langle e_{xy} \rangle = 0$. Therefore, the stabilization of an improper polarization $\langle p_y \rangle$ or $\langle p_z \rangle$ requires the stabilization of secondary spin waves $S_y$ or $S_z$ in quadrature with $S_x$. Notice that, in any case, the secondary spin waves are triggered by the primary order parameter and, consequently, share the same wave-vector. All these features are, in fact, observed experimentally.

By replacing (8) into (7), one can then express the free energy density as a function of the new set of magnetic order parameters $\sigma_x$, $\sigma_{yF}$, $\sigma_{yA}$, $\sigma_{zF}$ and $\sigma_{zA}$:

$$g(x) = \left[ t\cos^2(qx) - \frac{q^2}{4}\sin^2(qx) + \frac{q^4}{4}\cos^2(qx) \right]\sigma_x^2 + \frac{1}{4}\left[ \cos^4(qx) + 4\mu q^2 \sin^2(qx)\cos^2(qx) \right]\sigma_x^4 -$$

$$- \frac{1}{4}\left[ \sigma_{yF}^2 \cos^2(qx) + \sigma_{yA}^2 \sin^2(qx) \right]^2 - \frac{1}{4}\left[ \sigma_{zF}^2 \cos^2(qx) + \sigma_{zA}^2 \sin^2(qx) \right]^2 \qquad (9).$$

### D. The free energy of the competing phases

The general free energy functional $G$ can now be obtained by averaging (9) over one period of the magnetic modulation [ $G = \frac{1}{x_0}\int_0^{x_0} g(x)dx$ ]. This leads to:

$$G = \frac{1}{2}\left[ t - \frac{q^2}{4} + \frac{q^4}{4} \right]\sigma_x^2 + \frac{1}{4}\left[ \frac{3}{8} + \frac{\mu q^2}{2} \right]\sigma_x^4 - \frac{1}{4}\left[ \frac{3}{8}(\sigma_{yF}^4 + \sigma_{yA}^4) + \frac{1}{4}\sigma_{yA}^2\sigma_{yF}^2 \right] -$$
$$- \frac{1}{4}\left[ \frac{3}{8}(\sigma_{zF}^4 + \sigma_{zA}^4) + \frac{1}{4}\sigma_{zA}^2\sigma_{zF}^2 \right] \qquad (10),$$

Notice that in (10) the values of $\sigma_{yF}$, $\sigma_{yA}$, $\sigma_{zF}$ and $\sigma_{zA}$, given by (8), correspond to the their equilibrium values if $\sigma_x$ minimizes $G$. Therefore, the stability of the different competing magnetic phases can be simply determined by imposing the equilibrium condition $\frac{\partial G}{\partial \sigma_x} = 0$, together with the stability conditions $\frac{\partial^2 G}{\partial \sigma_i^2} > 0$ and $\det \left[ \dfrac{\partial^2 G}{\partial \sigma_i \partial \sigma_j} \right] > 0$. For simplicity, we will consider only the potential stability of the phases that are experimentally observed. Consequently, we will ignore mixed phases where two or more of the secondary parameters $\sigma_{yF}$, $\sigma_{yA}$, $\sigma_{zF}$ and $\sigma_{zA}$ co-exist, or ferroelastic phases with non-zero $\sigma_{zF}$ or $\sigma_{yF}$. We will focus on the competition between four relevant phases: the longitudinal incommensurate phase (L-INC), the cycloidal polar phases corresponding to the order parameters $\sigma_x$ and $\sigma_{yA}$ ($\vec{P}//\vec{b}$, Cycl-XY) or $\sigma_x$ and $\sigma_{zA}$ ($\vec{P}//\vec{c}$, Cycl-XZ) and the homogeneous antiferromagnetic (A-AFM) phase.

• PHASE-1: the AFM phase ($\sigma_x \neq 0$ and $\sigma_y = \sigma_z = 0$)

The free energy corresponding to the homogeneous ($\vec{q}=0$) anti-ferromagnetic phase is $G_1 = -t^2$ and the temperature and magnetic field dependences of the antiferromagnetic order parameter is $\sigma_x = \sqrt{-2t}$. This phase is potentially stable if $t < 0$.

• PHASE-2: the L-INC phase ($\sigma_x \neq 0$ and $\sigma_y = \sigma_z = 0$)

The free energy ($G_2$) and the amplitude of the magnetic modulation ($\sigma_x$) for this modulated phase ($\vec{q} \neq 0$) are given by:

$$G_2 = -\dfrac{1}{4} \dfrac{\left[ t - \dfrac{q^2}{4} + \dfrac{q^4}{4} \right]^2}{\left[ \dfrac{3}{8} + \dfrac{1}{2}\mu q^2 \right]} \tag{12a},$$

$$\sigma_x^2 = -\dfrac{\left[ t - \dfrac{q^2}{4} + \dfrac{q^4}{4} \right]}{\left[ \dfrac{3}{8} + \mu \dfrac{q^2}{2} \right]} \tag{12b}$$

This phase will be stable if $\left[t - \frac{q^2}{4} + \frac{q^4}{4}\right] < 0$ (that is, $\sigma_x^2 > 0$). The temperature dependence of the incommensurate modulation wave-vector, which can be obtained from the equilibrium condition $\frac{\partial G_2}{\partial q} = 0$, is given by the equation:

$$2q\left[t - \frac{q^2}{4} + \frac{q^4}{4}\right]\left\{\left(q^2 - \frac{1}{2}\right)\left(\frac{3}{8} + \mu\frac{q^2}{2}\right) - \frac{\mu}{2}\left(t - \frac{q^2}{4} + \frac{q^4}{4}\right)\right\} = 0. \tag{12c}$$

The solutions $q = 0$ and $\left[t - \frac{q^2}{4} + \frac{q^4}{4}\right] = 0$ correspond to the non-modulated antiferromagnetic phase and to the paramagnetic phase, respectively. The other solutions are

$$q_\pm^2 = \frac{1}{6\mu}\left\{(\mu - 3) \pm 3\sqrt{1 + \frac{\mu^2}{9}(1 + 48t) + \frac{4}{3}\mu}\right\}. \tag{12d}$$

As can be seen from (12c), at the second order transition point from the paramagnetic to the longitudinal incommensurate phase ($\sigma_x^2 \to 0$), these additional solutions correspond to $q_+^2 = \frac{1}{2}$ and $q_-^2 = -\frac{3f}{4\mu}$. Since $\mu$ can be a positive constant, this latter solution must be discarded and the former one identified as that corresponding to the incommensurate phase. Then, from the condition $G_2(t_i, b = 0) = 0$, one finds $t_i = \frac{1}{16}$. The experimental and reduced temperature and wave vector scales are therefore related as

$$t = \frac{(T - T_0)}{16(T_i - T_0)},$$
$$q(t) = \frac{1}{\sqrt{2}}\frac{k(T)}{k(T_i)} \tag{12e}$$

where $T_i$ represents the experimental temperature of the zero magnetic field transition between the paramagnetic and the incommensurate phases.

• PHASE-3: the Cycl-XY phase ($\sigma_x \neq 0$ and $\sigma_{yA} \neq 0; \sigma_{zA} = \sigma_{zF} = \sigma_{yF} = 0$)

If, in addition to the primary longitudinal modulation $\sigma_x \neq 0$ there exists a secondary modulation of the y-component of the Mn spins that is in quadrature with $\sigma_x$, then the phase will develop, as seen, a spontaneous electrical polarization along the b-axis given by $p_y = \nabla_2 q \sigma_x \sigma_{yA}$, while the ferroelastic deformation averages out. The energy of this ferroelectric phase is

$$G_3 = -\frac{1}{4} \frac{\left[t - \frac{M}{4}q^2 + \frac{1}{4}q^4\right]^2}{\left[\frac{3}{8} + \frac{\mu}{2}q^2 - Nq^4\right]} - \frac{3}{8}\frac{\left[(t-t_1)\right]^2}{4} \quad , \tag{13c}$$

where $M = 1 - \frac{3\nabla_2^2}{2}$ and $N = \frac{3\nabla_2^4}{8}$. The amplitude of the primary modulation and the stability conditions are respectively given by:

$$\sigma_x^2 = -\frac{\left[t - \frac{M}{4}q^2 + \frac{1}{4}q^4\right]}{\left[\frac{3}{8} + \frac{\mu}{2}q^2 - Nq^4\right]} \quad , \tag{13d}$$

and

$$\left[t - \frac{M}{4}q^2 + \frac{1}{4}q^4\right] < 0$$
$$\left[\frac{3}{8} + \frac{\mu}{2}q^2 - Nq^4\right] > 0 \tag{13e}$$
$$\nabla_2^2 q^2 \sigma_x^2 > a_y(t - t_1)$$

As before, the temperature dependence of the incommensurate modulation wave-vector can be obtained from the equilibrium condition $\frac{\partial G_3}{\partial q} = 0$. This leads to the equation:

$$2q\left[t - M\frac{q^2}{4} + \frac{q^4}{4}\right]\left\{\left(q^2 - \frac{M}{2}\right)\left(\frac{3}{8} + \mu\frac{q^2}{2} - Nq^4\right) - \left(\frac{\mu}{2} - 2Nq^2\right)\left(t - M\frac{q^2}{4} + \frac{q^4}{4}\right)\right\} = 0 \quad (13f)$$

and to a temperature dependence of the cycloidal wave vector given by one real root of the equation:

$$-\frac{N}{2}q^6 + \frac{3}{8}\mu q^4 + \left(\frac{3}{8} - \frac{\mu M}{8} - 2Nt\right)q^2 + \left(\frac{3}{16}M - \frac{\mu t}{2}\right) = 0 \quad (13g)$$

• PHASE 4: the Cycl-XZ phase ($\sigma_x \neq 0$ and $\sigma_{zA} \neq 0; \sigma_{zF} = \sigma_{yA} = \sigma_{yF} = 0$)

This case is similar to the previous one but with the cycloid lying on the $xz$ plane and the spontaneous polarization directed along the $z$-axis ($p_z = \xi_2 q \sigma_x \sigma_{zA}$). The energy of this phase is given by:

$$G_4 = -\frac{1}{4}\frac{\left[t - \frac{M'}{4}q^2 + \frac{1}{4}q^4\right]^2}{\left[\frac{3}{8} + \frac{\mu}{2}q^2 - N'q^4\right]} - \frac{3}{8}\frac{[(t - t_2)]^2}{4} \quad (14a),$$

where $M' = 1 - \frac{3\xi_2^2}{2}$ and $N' = \frac{3\xi_2^4}{8}$. The amplitude of the primary modulation and the stability conditions are respectively given by:

$$\sigma_x^2 = -\frac{\left[t - \frac{M'}{4}q^2 + \frac{1}{4}q^4\right]}{\left[\frac{3}{8} + \frac{\mu}{2}q^2 - N'q^4\right]} \quad (14b)$$

and

$$\left[t - \frac{M'}{4}q^2 + \frac{1}{4}q^4\right] < 0$$

$$\left[\frac{3}{8} + \frac{\mu}{2}q^2 - N'q^4\right] > 0 \quad (14c),$$

$$\xi_2^2 q^2 \sigma_x^2 > (t - t_2)$$

Finally, the equilibrium condition $\frac{\partial G_4}{\partial q} = 0$ leads to the equation

$$-\frac{N'}{2}q^6 + \frac{3}{8}\mu q^4 + \left(\frac{3}{8} - \frac{\mu M'}{8} - 2N't\right)q^2 + \left(\frac{3}{16}M' - \frac{\mu t}{2}\right) = 0, \quad (14d)$$

which can be solved in order to $q$ to give the temperature dependence of the incommensurate modulation wave-vector, common to both magnetic order parameters.

## 4. MODELLING THE PHASE DIAGRAMS OF THE RMnO$_3$ COMPOUNDS

As seen above, the crystalline distortion that leads to magnetic frustration and, eventually, to the ferroelectric order, mainly originates from the ionic radii of the rare-earth ions. This geometric effect can be controlled either in a stepwise manner, by using rare-earth elements with different ionic radius (from R=La$^{3+}$ to Pr$^{3+}$, Eu$^{3+}$, Gd$^{3+}$, Tb$^{3+}$, Dy$^{3+}$, Ho$^{3+}$, Yb$^{3+}$ and Lu$^{3+}$), or quasicontinuously, by tuning the average rare-earth radius in solid solutions in which the R$^{3+}$ ion is partially replaced by one isoelectric ion with different ionic radius, as in the case of the Eu$_{1-x}$Y$_x$MnO$_3$ mixed system.[52-55] In the following we will analyze these two cases within the scope of the model presented above, as illustrative examples.

### A. The pure compounds

Figure 2 shows the experimental temperature dependence of the modulation wave vector in the Gd, Tb, Dy and Ho compounds (dots; data taken from Ref. 20). As seen in the preceding section, the evolution of the modulation wave vector with temperature in the L-INC phase is solely determined, in the model, by the signal and magnitude of the coupling constant µ, along with the value of $T_0$. For a given compound, these two constants can be estimated by fitting $q_+(t,\mu)$ given by equation (12d) to the experimental $\delta_{inc}(T)$ curve, and by taking into account that the relationship between the experimental and the reduced temperature and wavelength scales is given by (12e). The curves fitted to the experimental data in this way (see lines in figure 2) reproduce well the essential features of the observed behaviour of the modulation wavenumber $\delta(T)$.

In the L-INC phase and on cooling, $\delta(T)$ decreases for Gd and Tb and increases for Dy and Ho. This behaviour implies a transition from a positive to a negative µ, as the ionic radius $R_{ion}$ decreases. As shown in figure-3a (see also table III), µ varies almost linearly with the ionic radius of the rare-earth elements, from µ~1 for Gd to µ~ -0.7 for Ho. The value µ=0, for which the modulation wavelength is independent of the temperature, can be estimated from the linear fit as R~105.7 pm, a value that is intermediate between Tb (106.3 pm) and Dy (105.2 pm). Given the linear relationship between $\delta(T_i)$ and $R_{ion}$, this value would correspond to the commensurate value $\delta(T_i) \sim 1/3$. These conclusions are in excellent agreement with the behaviour experimentally observed in the $Tb_{1-x} Dy_x MnO_3$, where a spin modulation with a temperature independent wave number $\delta \sim \frac{1}{3}$ is observed for compositions in the range 0.5<x<0.68 (Ref. 56).

The dependence of $T_0$ on the rare earth ionic radius is not very pronounced for the Gd, Tb, Dy and Ho compounds, varying only slightly within the range 32-36K (see figure 3b). It is interesting, however, to consider how this parameter varies outside this range of rare-earth elements. As seen, $T_0$ corresponds to the temperature for which the magnon branch softens at the centre of the Brillouin zone. For compounds such as $LaMnO_3$,

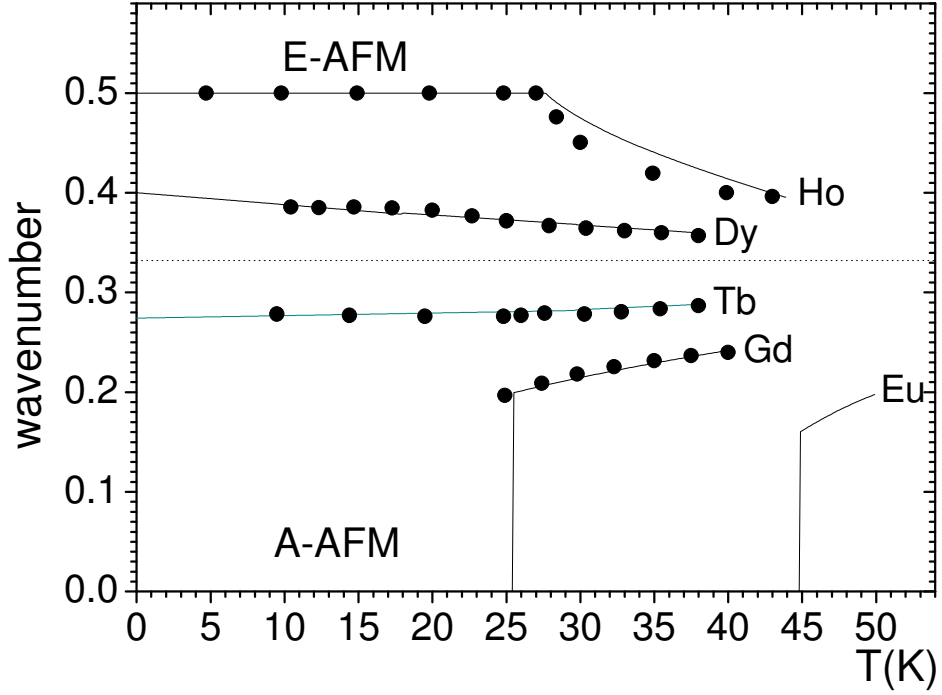

FIG. 2. (a) Experimental (dots) and simulated (lines) temperature dependences of the modulation wavenumbers for the RMnO$_3$ compounds (R=Gd, Tb,Dy and Ho). For EuMnO$_3$ only the simulation is shown (see text).

TABLE III. The values of the model parameters µ, $\nabla_2$, $T_0$ and $T_1$ that allow the simulation of the phase transition sequences observed experimentally. The values of $T_{inc}$ and $\delta(T_i)$ are also given.

|                     | µ     | $\nabla_2$ | $T_0$ | $T_1$ | $T_{inc}$ | $\delta(T_i)$ |
|---------------------|-------|------------|-------|-------|-----------|---------------|
| EuMnO$_3$           | 1.35  | --         | 48    | --    | 50        | 0.198         |
| GdMnO$_3$           | 1     | --         | 35    | --    | 40        | 0.242         |
| TbMnO$_3$           | 0.2   | 0.6        | 31.5  | 24    | 38        | 0.288         |
| DyMnO$_3$           | -0.15 | 0.3        | 33    | 15    | 38        | 0.36          |
| HoMnO$_3$           | -0.68 | --         | 35    | --    | 44        | 0.395         |

PrMnO$_3$, NdMnO$_3$ and SmMnO$_3$, for which frustration does not occur, $T_0$ corresponds to the critical temperature of the direct transition between the PM and the A-AFM phases. This implies that $T_0$ decreases rather steeply as $R_{ion}$ decreases in the range 107 pm< $R_{ion}$ <117 pm (see inset of figure 2b), stabilizing at a more or less constant values once magnetic frustration is achieved and the intermediate L-INC phase is induced (within the range from Eu to Ho).

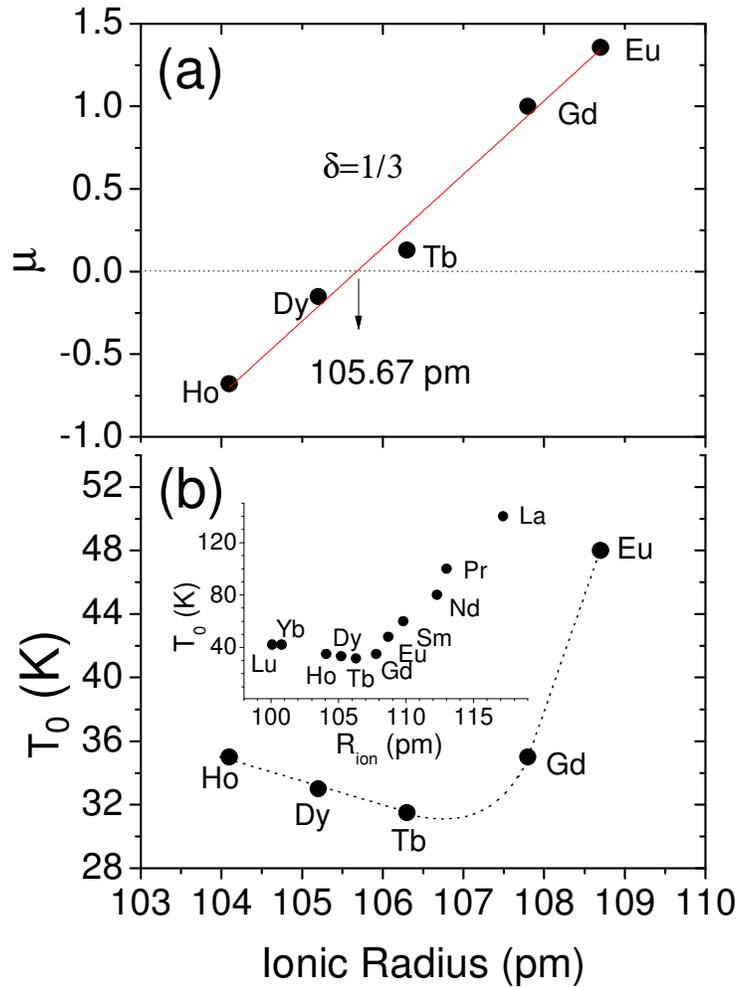

FIG. 3. The dependence of the parameters (a) μ and (b) $T_0$ on the ionic radius of the rare-earth ion R (from Eu to Ho). The inset in (b) shows the variation of $T_0$ over an extended range that includes R= La, Sm Pr , Nd, Yb and Lu (see text).

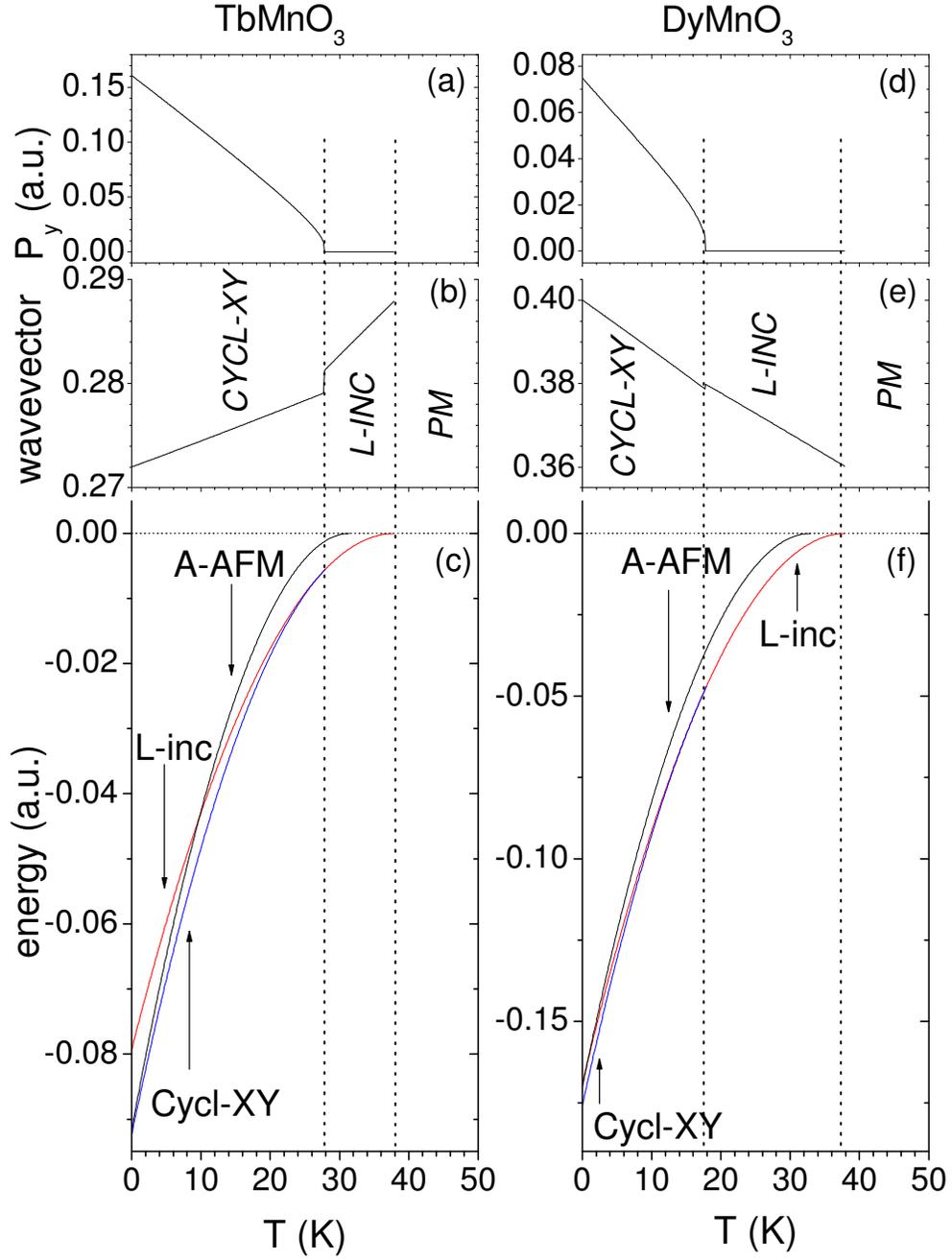

FIG. 4. Polarization, modulation wavevector and free energy of the phases L-INC, Cycl-XY and A-AFM as functions of temperature for TbMnO$_3$ (a-c) and DyMnO$_3$ (d-f). The model parameters are given in table III and the reference energy corresponds to that of the paramagnetic phase.

For the case of EuMnO$_3$ there are no reported data on the temperature dependence of $\delta(T)$ within the narrow temperature range of stability of the L-INC phase (50K<T<46K). However, from the linear dependence of µ and $\delta(T_i)$ on $R_{ion}$, one can estimate, for this compound ($R_{ion} = 108.7\,pm$), the values µ=1.36 and $\delta(T_i) \sim 0.19$. Then, if we adopt these values, we are left with a single parameter ($T_0$) to fit the critical temperature of the transition between the L-INC and the A-AFM phases. The fit of this unique parameter gives $T_0 \sim 48K$, a value that is entirely consistent with the general trend of $T_0(R_{ion})$ seen in the inset of figure 3. In addition, the set of parameters thus found for the Eu compound (µ, $T_0$ and $\delta(T_i)$) allows us to estimate the function $\delta(T)$ within the L-INC phase. This estimated temperature dependence of the modulation wave number is depicted in figure 2.

For the cases of the Tb and the Dy systems, the ground state is the cycloidal modulated phase (Cycl-XY), which is polar ($\vec{P}//\vec{b}$). Here, we have to include the analysis of the potential stability of this additional phase and tune the additional parameters $\nabla_2$ and $T_1$ in order to account for both the observed transition temperature from the L-INC to the Cycl-XY phase and the temperature dependence of the modulation wave vector in the range of stability of this lower temperature phase. As seen in figure 4, the values given in table III for these additional parameters allow the simulation of the phase sequence PM→L-INC→ Cycl-XY observed in these compounds at zero magnetic field. It is also possible to calculate the temperature dependence of both the electric polarization and the modulation wave vector near the transition from the L-INC to the Cycl-XY phase. These quantities are also plotted in figure 4 for these two intermediate compounds.

### B. The solid solutions: the example of $Eu_{1-x}Y_x MnO_3$

The model can also be applied to the description of the phase diagrams of solid solutions in which the average value of the radius of the rare-earth element is tuned by the partial substitution of isoelectric ions. Here, we will analyse, as one illustrative example, the case of the $Eu_{1-x}Y_x MnO_3$ mixed system.

In the $Eu_{1-x}Y_xMnO_3$ solid solution, the *Pnma* orthorhombic symmetry is maintained only for *x<0.6*. Above this concentration, traces of the *P63cm* hexagonal phase of $YMnO_3$ appear. As $x$ increases, the volume and the orthorhombic distortion of the unit cell cross the values found in $GdMnO_3$ (x~0.2) and $TbMnO_3$ (x~0.8). Despite the continuous shrinking of the lattice volume, the in-plane orthorhombic distortion, parameterized by $\varepsilon = (a-c)/(c+a)$, tends to saturate near $x$~0.4 (Ref. 14). Notice that the inequality of the lattice constants $a$ and $c$ reflects the tilting of the oxygen octahedral around the b-axis and the consequent reduction of Mn-O-Mn bond angle. It is likely that this reduction may not be the only factor affecting the spin system. The shrinkage of the unit cell and the A-site disorder may also affect the orbital overlap and the magnetic exchange.

In the composition range 0<$x$<0.5, the ($x$-$T$) diagram of the solid solution has been investigated by different groups.[14,52,54] The temperature range of stability of the L-INC phase increases from 51K>T>46K at $x$=0 to 45K>T>22K at $x$=0.5. For $x$<0.2, the A-AFM phase is stabilized at low temperatures, although within a temperature band that rapidly narrows as $x$ decreases. For $x$>0.3 the A-AFM phase is suppressed and the low temperature phase corresponds to a P//z ferroelectric phase. For $x$=0.2 the canted ferromagnetism (M//y) characteristic of the A-AFM co-exists, at low temperatures, with the P//z polarization.[14] This co-existence may signal either the stabilization of a more complex magnetic phase[14] or a co-existence of both phases. Although there is no direct experimental evidence, the P//z phase is attributed to the stabilization of the Cycl-XZ phase,[52] as observed in $GdMnO_3$ under a magnetic field.

In principle, for a solid solution, one can find the adequate model parameters by following the procedure described above for the case of the pure systems. That is, one can fit, for each composition, the transition temperatures and the temperature dependences of the modulation wave vectors within the ranges of stability of the different phases observed. Quite often, complete experimental information is not available but, for $Eu_{1-x}Y_x MnO_3$, it exists at least in part.

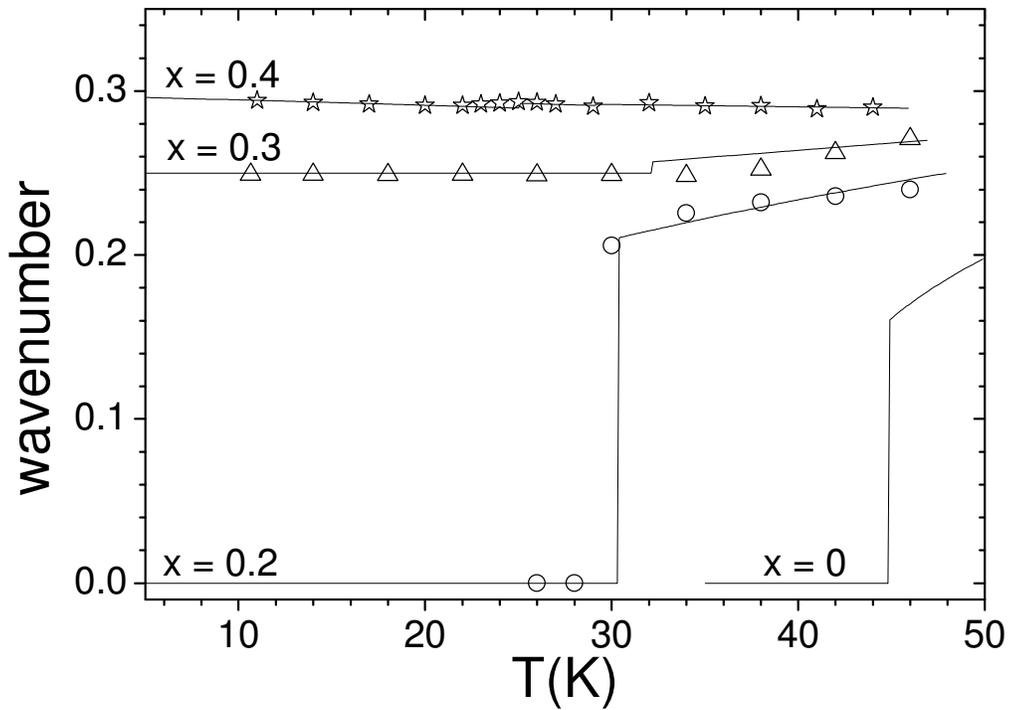

FIG. 5. Experimental (dots) and simulated (lines) of the temperature dependence of the wave numbers of the magnetic modulation in the $Eu_{1-x}Y_x MnO_3$ mixed system for the compositions $x=0.2$, $x=0.3$ and $x=0.4$. The experimental data was taken from Ref. 54. The predicted modulation wave number for pure $EuMnO_3$ is also shown for comparison.

TABLE IV. The values of the model parameters µ, $\nabla_2$, $\xi_2$ $T_0$, $T_1$ and $T_2$ allowing the simulation of the (T-x) dependence of the modulation wave vector and of the phase transition sequences observed experimentally in the three compositions. The experimental values of $T_{inc}$ and $\delta(T_i)$ are also given. The $\beta_{eff}(1/4)$ coefficient corresponds to the effective lock-in potential to the C-phase with δ=1/4.

| | µ | $\nabla_2$ | $\xi_2$ | $T_0$ | $T_1$ | $T_2$ | $T_{inc}$ | $\delta(T_i)$ | $\beta_{eff}(1/4)$ |
|---|---|---|---|---|---|---|---|---|---|
| $EuMnO_3$ | 1.35 | -- | -- | 48 | -- | -- | 50 | 0.198 | 0.003 |
| $Eu_{0.8}Y_{0.2}MnO_3$ | 1.04 | 0.5 | 0.7 | 38.68 | 23.5 | 25.85 | 48 | 0.243 | 0.003 |
| $Eu_{0.7}Y_{0.3}MnO_3$ | 0.59 | 0.5 | 0.7 | 35.47 | 23.5 | 23.03 | 47 | 0.266 | 0.003 |
| $Eu_{0.6}Y_{0.4}MnO_3$ | -0.04 | 0.5 | 0.7 | 33.26 | 23.5 | 17 | 46 | 0.290 | 0.003 |

The experimental temperature dependence of the magnetic modulation wave number $\delta(T)$ is shown in figure-5 for the compositions x=0.2, x=0.3 and x=0.4 (Ref. 54). The curve predicted for $EuMnO_3$ (see above) is also shown for comparison. For a given composition, the model parameters can be fit in order to simultaneously reproduce $\delta(x,T)$ and the temperature ranges of stability of the phases observed. For the three compositions shown, the values of the model parameters obtained in this way are listed in table IV. Notice that the parameters µ and $T_0$ can be determined for the three compositions because the L-INC phase is always stable. However, for x=0.3, the wavenumber corresponding to the Cycl-XZ phase locks at the commensurate value $\delta = \frac{1}{4}$. Therefore, only in the case of x=0.4, where both the incommensurate Cycl-XY and Cycl-XZ are stable, can one explicitly determine the values of $\nabla_2$, $\xi_2$, $T_1$ and $T_2$. This has been done by carefully fitting, for this composition, the temperature ranges of stability of the two cycloidal phases and the experimental temperature dependence of the modulation wave number (see figure-6a). The parameters adjusted is this way allow us to calculate the temperature dependence of the electrical polarizations $P_y$ and $P_z$ and to describe the polarization rotation observed at the transition between the Cycl-XY and the Cycl-XZ phases (figure 6b). Due to the limited experimental information, we have decided to maintain, for the other compositions, the values of $\nabla_2$, $\xi_2$ and $T_1$ as determined for $x=0.4$.

As seen above, the cycloidal-XZ spin modulation observed for $x=0.3$ is commensurate ($\delta=1/4$). In an incommensurate phase, we consider the modulation wave number as a variational parameter whose value, at equilibrium, results from the condition $\dfrac{\partial G}{\partial q}=0$.

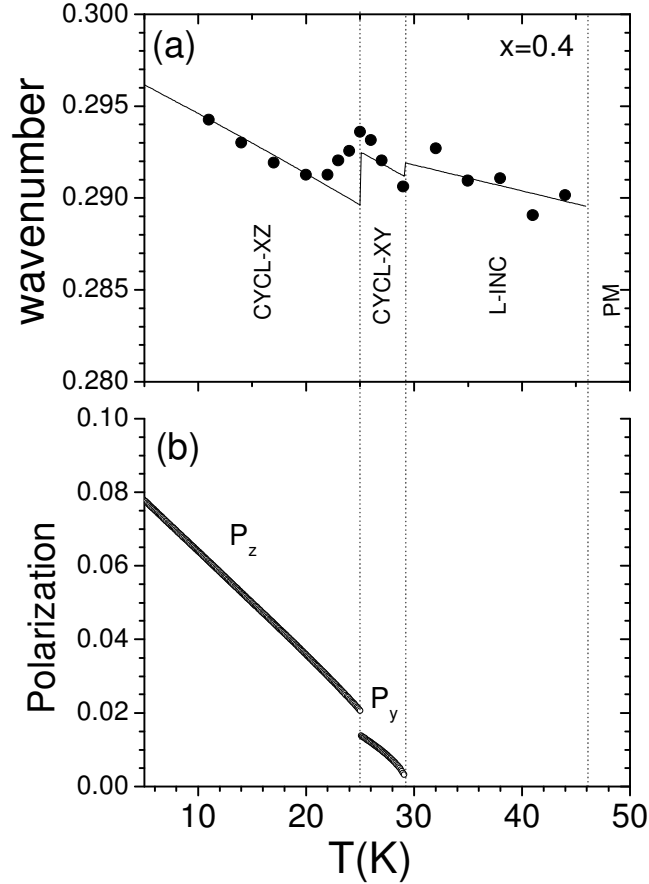

FIG. 6. (a) Detail of the experimental (dots) and fitted (line) temperature dependence of the modulation wave number of the spin wave for $x=0.4$;
(b) The electric polarizations $P_y$ and $P_z$ as functions of the temperature. The polarization rotation observed at the transition between the Cycl-XY and the Cycl-XZ phases is well reproduced by the model.

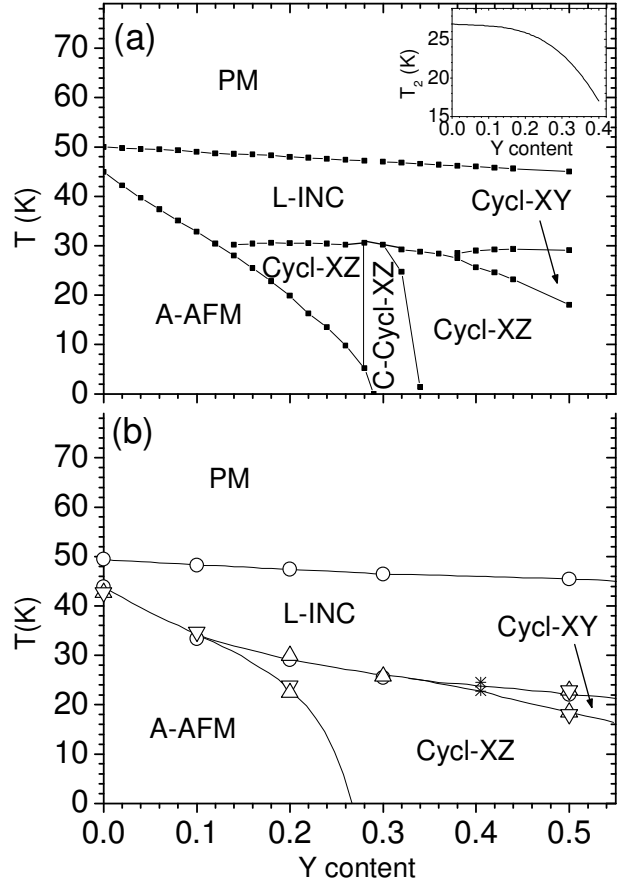

FIG. 7. Simulated (a) and experimental (b) temperature versus composition phase diagram of the $Eu_{1-x}Y_x MnO_3$ solid solution. The experimental phase diagram was taken from Ref. 14.

However, a given commensurate phase has a fixed rational modulation wave-number $\delta = \frac{n}{p}$. In general, this fixed value of $\delta$ costs energy, when compared to the incommensurate solution considered above. This cost may be compensated by the additional terms (Umklapp invariants) that are, in this case, allowed by symmetry. These Umklapp terms are of degree $2p$ ($p>2$) for a commensurate wave-number $\delta = \frac{n}{p}$ ($n$ and $p$ integers). In the case of the $\delta=1/4$ cycloidal phase observed for $x=0.3$, we have assumed, for simplicity, that the effect of the lock-in potential $U_{umklapp} = -\beta \sigma_x^8$ could be described by means of effective fourth degree terms in the primary order parameter amplitude ($U_{umklapp} \approx \frac{1}{4}\beta_{eff}\sigma_x^4$). This approximation can be justified by the fact that the δ=1/4 phase is observed in a temperature range well below $T_i$, implying that the temperature dependence of the amplitude of the order parameter $\sigma_x$ is already weak ($\sigma_x \sim (T-T_i)^{1/2}$). Consequently, the temperature dependence of $\beta_{eff}$ is not important and can be neglected. In other words, it is not the degree of the lock in term but rather its presence and magnitude that modify the free energy. In this approximation, the energy of the δ=1/4 phase can be estimated by replacing in the incommensurate free energy $q(t) = \frac{1}{\sqrt{2}}\frac{\delta(T)}{\delta(T_i)}$ by $q_c = \frac{1}{\sqrt{2}}\frac{1/4}{\delta(T_i)}$ and by adding the contribution of the effective lock-in potential averaged over a period of the modulation wave, $U = -\frac{3}{8}(\frac{1}{4}\beta_{eff}\sigma_x^4)$. For $x$=0.3, we have used in the simulation the minimum value of $\beta_{eff}$ required to stabilize the commensurate phase and the value of $T_2$ necessary to fit the observed transition temperature ($\beta_{eff} = 0.003$). We have maintained this value of $\beta_{eff}$ to estimate the free energy of the δ=1/4 phase for all the other compositions.

Under these circumstances, for any given value of $x$ within the range 0<$x$<0.5, the values of the parameters μ, $T_0$, $T_{inc}$ and $\delta(T_i)$ can be interpolated by polynomial fitting. By doing so, we maintain a single adjustable parameter ($T_2$) to model the observed (T,$x$) phase diagram of the solid solution. Notice that the value of $T_2$ solely influences the range of stability of the Cycl-XZ phase and the temperature dependence of the modulation wave number within this phase. As seen in figure 7, the experimental phase diagram taken from Ref. 14 can be very well reproduced if one assumes that $T_2$ decreases smoothly with $x$ as illustrated in the inset of the figure.

## 5. CONCLUDING REMARKS

The present paper described one unified Landau model for the phase diagrams of the rare-earth orthomanganese compounds. The common symmetry $\Gamma(B_2)$ of the primary magnetic order parameters over the whole set of the orthorhombic RMnO$_3$ compounds was stressed and used to obtain, in the simplest possible terms, an unified phenomenological description of the phase transition sequences observed. Besides the primary parameter, the model also includes two additional magnetic modes of symmetry $\Gamma(A_2)$ and $\Gamma(A_1)$, which couple bi-quadratically to the primary mode. This set of three active magnetic modes of distinct symmetries can be related to the softening of the three spatial components of the $\vec{A}$ eigen mode of the Mn$^{3+}$ spins. As in the case of typical displacive modulated systems, the type-II Landau description of modulated phases was used to generate adequate free energy functionals for the different competing phases.

The model is rooted in exact symmetry considerations. This fact guarantees the consistency of the overall picture and elucidates the possible coupling between the different degrees of freedom. In particular, the potential ferroelectric and ferroelastic properties of a given magnetic phase can be clearly established, either by searching for the allowed mixed invariants that are linear on a electrical polarization or a lattice deformation, or by establishing, directly, the magnetic symmetry of a given ordered phase.

The model is capable of generating the observed phase diagrams and account for the polar properties of a given compound or solid solution. However, magneto-electric biferroicity and ferromagnetoelectricity (the linear magneto-electric effect) are here entirely excluded by symmetry. That is, the phases considered in the present model (and observed experimentally in these compounds at zero magnetic field) may be improper ferroelectric phases but are not multi-ferroic phases, at least if the standard notion of ferroic order is adopted. Notice, for example, that the phase transitions observed in TbMnO$_3$ corresponds, on cooling, to the sequence $(Pnma)' \to P_a(P_{\bar{1}1S}^{nma}) \to P_a(P_{\bar{1}1S}^{n2_1a})$. This last phase, $P_a(P_{\bar{1}1S}^{n2_1a})$, is ferroelectric ($\vec{P}//\vec{b}$) but it is neither ferromagnetoelectric nor multiferroic. Also, the thermally induced polarization rotation observed in Eu$_{0.5}$Y$_{0.5}$MnO$_3$ corresponds to a phase transition involving the cycloidal phases Cycl-XY $[(P_a(P_{\bar{1}SS}^{n2_1a}), \vec{P}//\vec{b}]$ and Cycl-XZ $[(P_a(P_{\bar{1}SS}^{nm2_1}), \vec{P}//\vec{c}]$ which, once again, have symmetries that are incompatible with ferromagnetoelectricity or multiferroicity.